\documentclass[twocolumn,showpacs,preprintnumbers,amsmath,amssymb,letterpaper]{revtex4}
\usepackage{epstopdf}
\usepackage{graphicx}% Include figure files
\usepackage{dcolumn}% Align table columns on decimal point
\usepackage{bm}% bold math
\usepackage{color}
\begin{document}
\title{Direct Search for Low Mass Dark Matter Particles  with CCDs}

\author{J. Barreto$^1$, H. Cease$^2$, H.T. Diehl$^2$, J. Estrada$^2$,  B. Flaugher$^2$, N. Harrison$^2$ J. Jones$^2$, B. Kilminster$^2$, J. Molina$^{3}$, J. Smith $^{2}$ ,  T. Schwarz$^4$ and A. Sonnenschein$^2$}
%\author{lots of people}
\address{ $^{1}$Universidade Federal do Rio de Janeiro (UFRJ), Rio de Janeiro, Brazil  \\
	         $^{2}$Fermi National Accelerator Laboratory, Batavia, Illinois, USA \\
  	         $^{3}$Facultad de Ingenieria, Universidad Nacional de Asuncion (FIUNA), Asuncion, Paraguay\\
	         $^{4}$University of California at Davis, USA.\\
                      }

\date{\today}

\begin{abstract}

A direct dark matter search is performed using fully-depleted high-resistivity CCD detectors.  
Due to their low electronic readout noise (R.M.S. $\sim$7 eV) 
these devices operate with  a very low  detection threshold of 40 eV, making the 
search for dark matter particles with low masses ($\sim$ 5 GeV) possible. 
The results of an engineering run  performed in a shallow underground site
are presented, demonstrating the  potential of this technology in the low mass region.

\end{abstract}

\maketitle

\section{Introduction}

There have been several direct-detection experiments searching for dark matter (DM) performed in recent years, and several more in development \cite{exps}. 
Most of these experiments have been optimized for detecting the elastic scattering of DM particles with masses larger than 50 GeV, concentrating on the most natural region of the Weakly Interactive Massive Particle (WIMP) parameter space such that DM corresponds to the lightest  supersymmetric particle \cite{susy_DM}. Detection thresholds of a few keV are typical for such high mass DM searches. Recent 
results from experiments that operate with a lower threshold \cite{damalibra,cogent} have presented hints of a DM signal 
at low energies, the most notable being DAMA/Libra which claims a high-significance detection of an annual modulation
\cite{damalibra}. More recently, the CoGeNT Collaboration has seen a possible hint for a low mass DM signal using a
low threshold Ge detector \cite{cogent}. Motivated by these results, new theoretical interpretations have been developed for which
low thresholds are needed to directly detect DM \cite{bottino1, bottino2, bottino3, light_DM1,light_DM2,light_DM3,light_DM4}. In these models the DM particles are either much lighter than 
50 GeV or present a non-elastic interaction with the detector nuclei \cite{inelastic}.  Because of noise in detector electronics and efficiency issues, 
most experiments are not capable of significantly lowering the detection threshold to probe the new models. 
For this reason there is an effort by the experimental community to develop techniques 
that could detect low energy signals from DM \cite{cogent} \cite{juancollar}. This work
presents a DM search using Charge-Coupled Devices (CCDs) using a detection threshold of  ~40 eV electron equivalent energy (eVee).

CCDs used for astronomical imaging and spectroscopy commonly achieve readout 
noise levels of 2 e$^-$ R.M.S., equivalent to 7.2 eV of ionizing energy in Silicon. Though this
allows for a very low detection threshold, these detectors have not been considered for DM
searches because of their very low fiducial mass.
The recent development of thick, fully-depleted CCDs, ten times more massive than conventional CCDs,
has  changed the situation. The Dark Matter in CCDs (DAMIC) experiment is the first
DM search  to exploit this technology.

%\begin{figure}
%\begin{center}
%%\includegraphics[width=.8\columnwidth]{seeing.pdf}
%%\plotone{pixel-hc.pdf}
%\includegraphics[width=1.\columnwidth]{limits_13440.pdf}
%\includegraphics[width=1.\columnwidth]{limits_13440_legend.pdf}
%\end{center}
%\caption{Examples of cross section limits from existing and planned direct dark matter searches. The
%limits are very weak for WIMP mass below 10 GeV. }
%\label{fig:DM_limits}
%\end{figure}

\section{High resistivity CCD detectors}

In an effort to increase the near-IR photon detection efficiency of
CCDs, a group at Lawrence Berkeley National Laboratory (LBNL) 
developed detectors with a depletion region up to 300 $\mu$m thick \cite{LBNL1}. 
These devices are fabricated with high resistivity silicon
($\sim 10 {\rm k}\Omega^{.}$cm). Due to their higher efficiencies for detecting photons
with red and near infrared wavelengths, these devices have been selected
for astronomical instrumentation  \cite{SNAP, Hypersuprime}, in
particular, the Dark Energy Camera (DECam)  \cite{DECam, DES}.  The DAMIC experiment uses
DECam CCDs to search for DM.  

%A cartoon of the LBNL CCDs used for the DECam project
%is presented in Fig.(\ref{fig:DESCCD}).  

The DECam CCDs \cite{DECam, DECam_CCDs, DECam_CCDtest, SPIE_diehl, SPIE_estrada}
are  back-illuminated,  p-channel CCDs thinned to 250 $\mu$m 
and biased with 40 V  from the back side to achieve full depletion.
The positively-charged holes produced in the depletion region are stored in 
the buried channels, a few $\mu$m away from the the gate electrodes. 
Charge produced near the back surface must travel the full thickness of the device to reach the 
potential well. During this transit inside the depletion region, 
a hole could also move in the direction perpendicular to the pixel boundaries. 
This effect, called charge diffusion,  must be regulated to avoid a significant 
degradation in image quality. The CCDs used in most astronomical instruments 
are thinned to $< $40 $\mu$m to reduce charge diffusion. For the DECam CCDs,  
a substrate voltage is applied to the back surface to control diffusion and obtain 
acceptable image quality.

%\begin{figure}
%\begin{center}
%\includegraphics[width=.8\columnwidth]{fig1.pdf}
%\end{center}
%\caption{Schematic of  a DECam detector. It is a back-illuminated  250 $\mu$m thick,  p-channel CCD.
%For more details see Ref. \cite{LBNL1}}
%\label{fig:DESCCD}
%\end{figure}

%The main features that make the DECam CCDs good candidates for a
%direct dark matter search are: their low electronic readout noise and  their thickness. 

The two main features that make the DECam CCDs good candidates 
for a direct dark matter search are their thickness, which allows the CCD 
to have significant mass, and the low electronic readout noise, which permits 
a very low energy threshold for the ionization signal produced by nuclear recoils. 

DAMIC uses a single rectangular CCD readout with a Monsoon controller \cite{monsoon}.
The detectors have 4.2 million pixels with dimensions 15 $\mu$m x 15  $\mu$m (active mass
of 0.5 g), and are read by two amplifiers in parallel.
The detectors have an output stage with  an electronic gain of $\sim 2.5$ $\mu$V/e. 
The signal is digitized after correlated double sampling (CDS) and the noise performance
depends on the readout speed. The best noise, $\sigma <$  2e$^-$ (R.M.S.) per pixel,
was achieved with readout times of 50 $\mu$s per pixel  
\cite{DECam_CCDs, DECam_CCDtest, SPIE_diehl, SPIE_estrada}.

\begin{figure}
\begin{center}
\includegraphics[width=1\columnwidth]{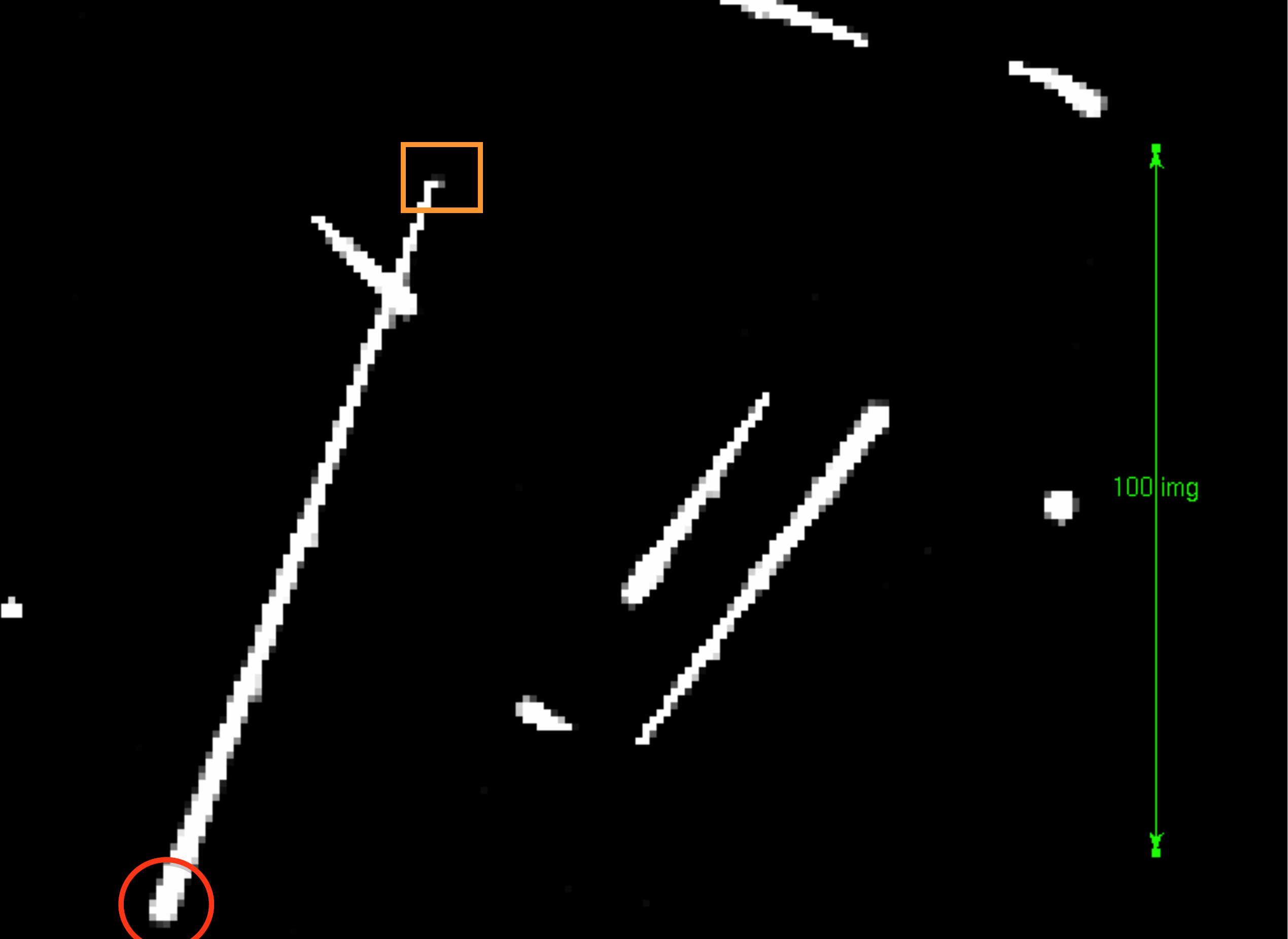}
\end{center}
\caption{Four muons tracks are clearly reconstructed on this image. There is a 
significant width different between the two ends of the track. The thicker end corresponds
to the region of the track in the back of the CCD, with maximum diffusion. The thinner
end corresponds to the front of the CCD. The smaller circular hits are diffusion-limited
hits as expected for nuclear recoils. The line is 100 pixels long and was  added to give a sense of scale.}
\label{fig:muon}
\end{figure}

\section{CCD Calibration: X-rays and nuclear recoils}

\begin{figure}
\begin{center}
\includegraphics[width=1.0\columnwidth]{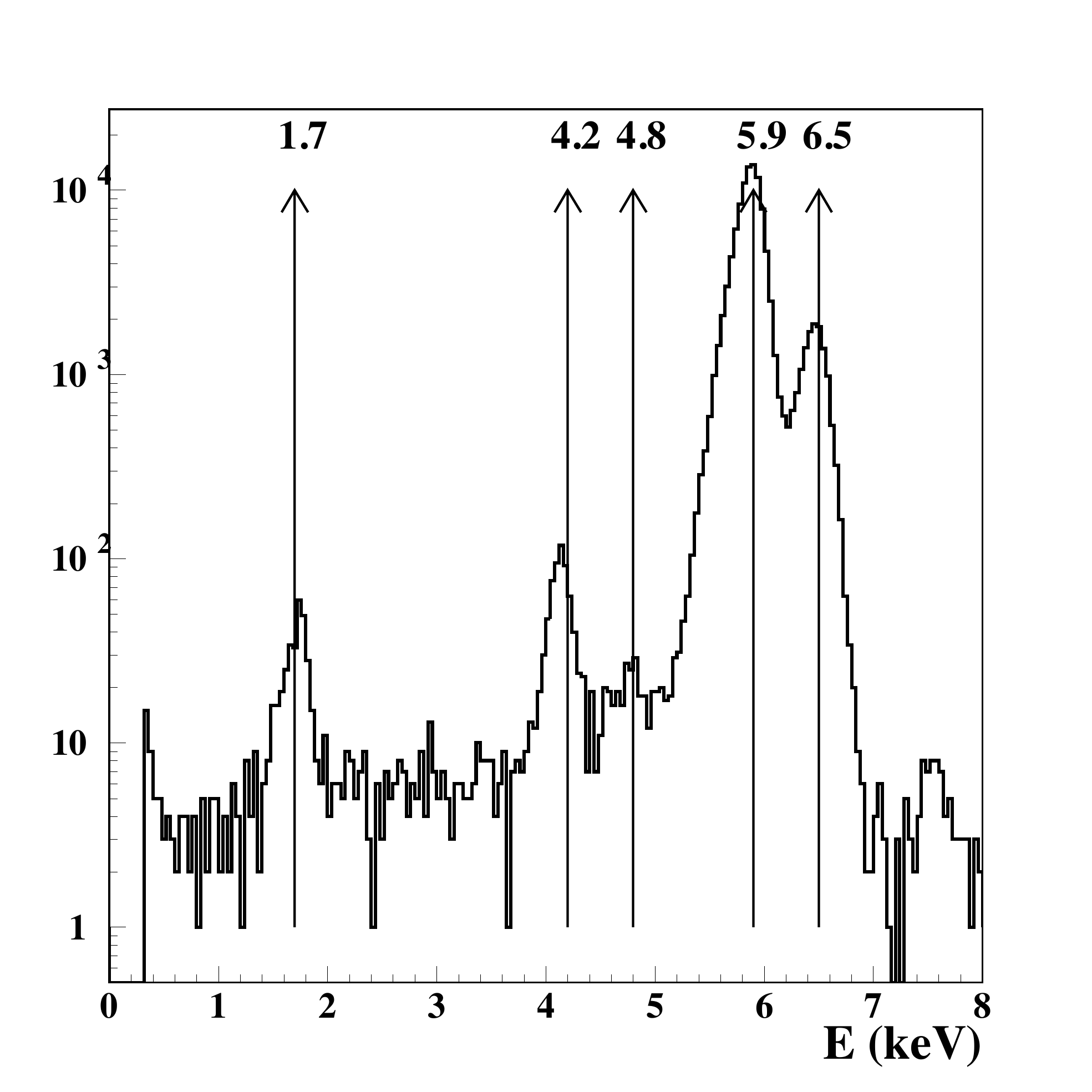}
\end{center}
\caption{Spectrum obtained for the reconstructed X-ray hits in an 
$^{55}$Fe exposure of a DECam CCD.  The arrows mark
the direct X-rays from the source: K$\alpha$=5.9 keV and K$\beta$=6.5 keV,
their escape lines at 4.2 and 4.8 keV \cite{janesick}, and
the Si X-ray at 1.7 keV.  The  factor 3.64 eV/e is used to convert from charge to ionization energy. 
The feature at 7.6 keV is consistent with pixels that hit by a 5.9 keV and a 1.7 keV
X-ray on the same exposure.
consistent w }
\label{fig:xrays}
\end{figure}

X-rays are commonly used 
in the energy calibration of CCD detectors \cite{janesick}. 
X-rays from $^{55}$Fe penetrate only $\sim$20 $\mu$m into the silicon before producing
charge pairs  according to the well known conversion
factor of 3.64 eV/e- \cite{janesick}.  As a result of this process X-rays produce hits 
in the detector with  size limited by charge diffusion.
In a back illuminated DECam CCD the size of the 5.9 keV X-rays is $\sim 7.5$ $\mu$m R.M.S. \cite{DECAM_diff}.
Using processing tools developed for astronomical aplications \cite{sextractor}, 
the X-ray hits are identified in the CCD images
as reconstructed charge clusters with a size limited by diffusion. 
Fig. \ref{fig:muon} shows two examples of diffusion-limited hits.
The energy of the X-ray hit is proportional to the charge collected in all
the pixels that belong to this cluster. An example of the 
energy spectrum measured for an  $^{55}$Fe  X-ray exposure in a 
DECam CCD is shown in Fig. \ref{fig:xrays}. 

%The diffusion measurement for DECam CCDs 
%using X-rays is shown in Fig. \ref{fig:xray_diff}. Similar measurements
%are done with other techniques in  Refs. \cite{DECAM_diff, LBNL_diff, LBNL_diff2}. 

%\begin{figure}
%\begin{center}
%\includegraphics[width=1.0\columnwidth]{noiseslow.pdf}
%\end{center}
%\caption{Noise as a function of pixel readout time for DECam CCDs. At slow readout speeds
%a noise below $\sigma = 2$ e is achieved. }
%\label{fig:noise}
%\end{figure}

%\begin{figure}
%\begin{center}
%\includegraphics[width=1.0\columnwidth]{xray_diff.pdf}
%\end{center}
%\caption{Diffusion measurement using X-rays from an  $^{55}$Fe source in a  250 $\mu$m DECam CCD with
%pixel size 15 $\mu$m x 15 $\mu$m. }
%\label{fig:xray_diff}
%\end{figure}

\begin{figure}
\begin{center}
\includegraphics[width=1.\columnwidth]{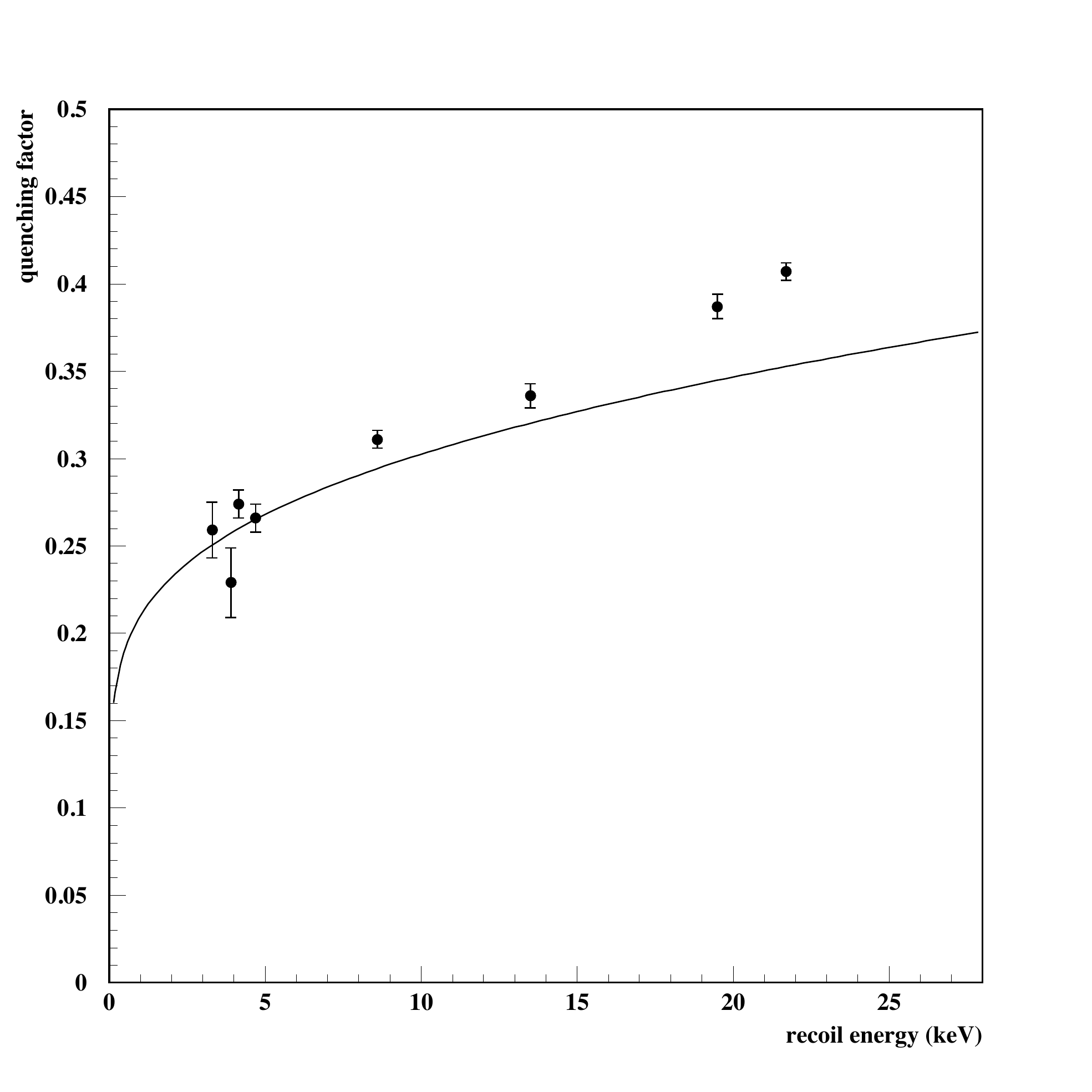}
\end{center}
\caption{Existing quenching factor measurements in Si compared with the Linhard theory (solid line). No previous data
exist for recoil energies below 4 keV\cite{Lindhard,Chagani}.}
\label{fig:quenching}
\end{figure}

Figure \ref{fig:xrays} provides a good calibration for 
X-rays in silicon.  Since the ionization efficiency for nuclear recoils 
differs from the ionization efficiency for X-rays, the results are not directly applicable to nuclear recoils
expected from DM interactions.  
%A significant fraction of the nuclear recoil 
%energy  is lost to acoustic vibrations or thermal excitations and does not produce charge 
%in the CCD. 
The ratio between the ionization efficiency for nuclear recoils and ionization efficiency for
electron recoils is usually referred to as the quenching factor ($Q$).
The quenching factor has been measured in Si  for 
recoil energies above 4 keV \cite{Lewin}, showing good agreement
with the Lindhard model  \cite{Lindhard, Chagani}. 
For recoil energies less than 4 keV, the quenching factor
becomes increasingly energy-dependent and  
no previous measurements are available, as shown in Fig.\ref{fig:quenching}.  
In this work, the Lindhard model  \cite{Lindhard, Chagani} is used
for converting visible (or electron equivalent) energy  to
recoil energy.   We test this assumption by comparison to neutron data with a known 
energy spectrum.

\begin{figure}
\begin{center}
\includegraphics[width=1.\columnwidth]{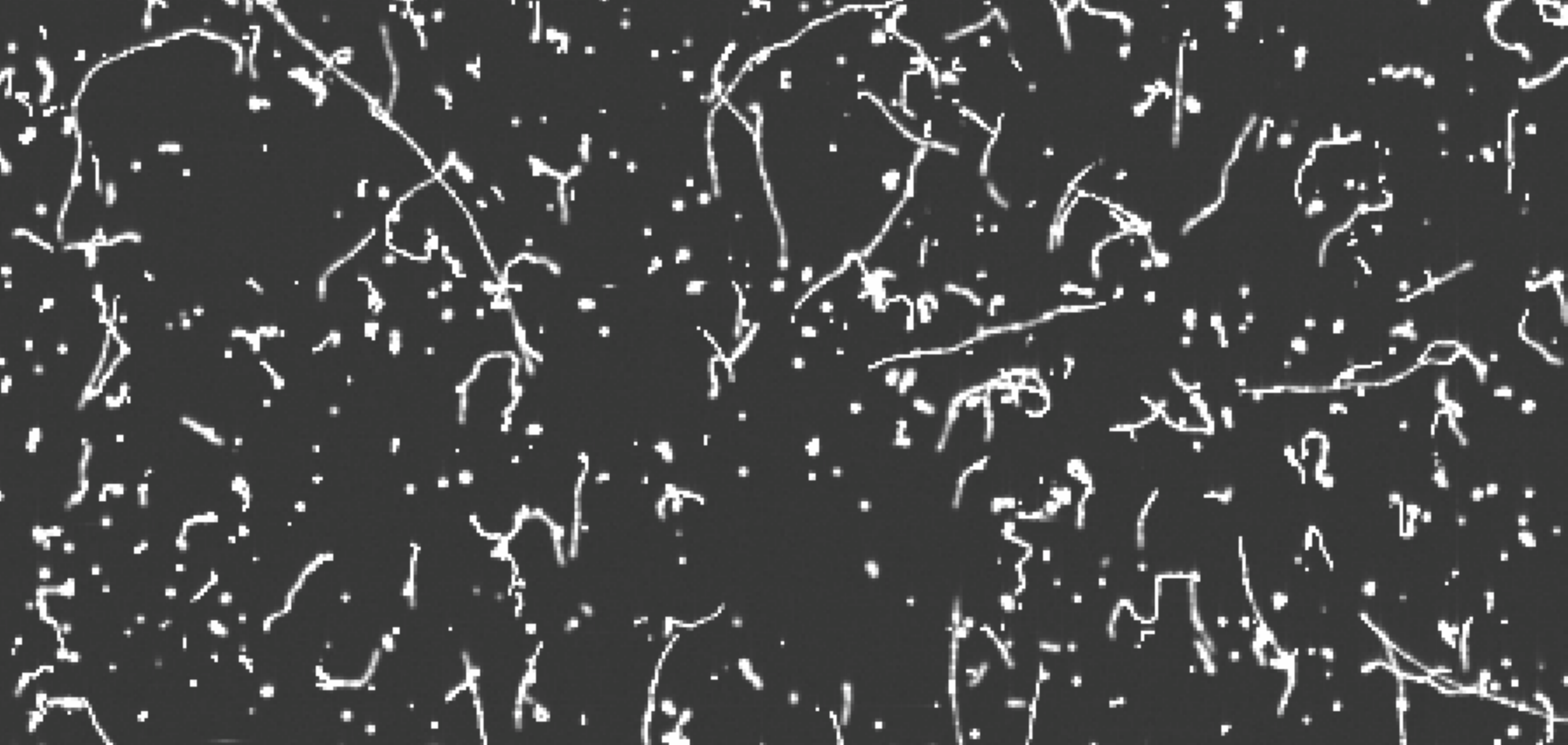}
\end{center}
\caption{Image resulting from an exposure of a DECam CCD to a $^{252}$Cf neutron source.
The total width of the image corresponds to 1000 pixels.  The smaller dots represent
the diffusion-limited hits, the trails correspond to scattered electrons and there is
one bigger circular cluster of charge that corresponds to an alpha particle. }
\label{fig:source}
\end{figure}

\begin{figure}
\begin{center}
\includegraphics[width=1\columnwidth]{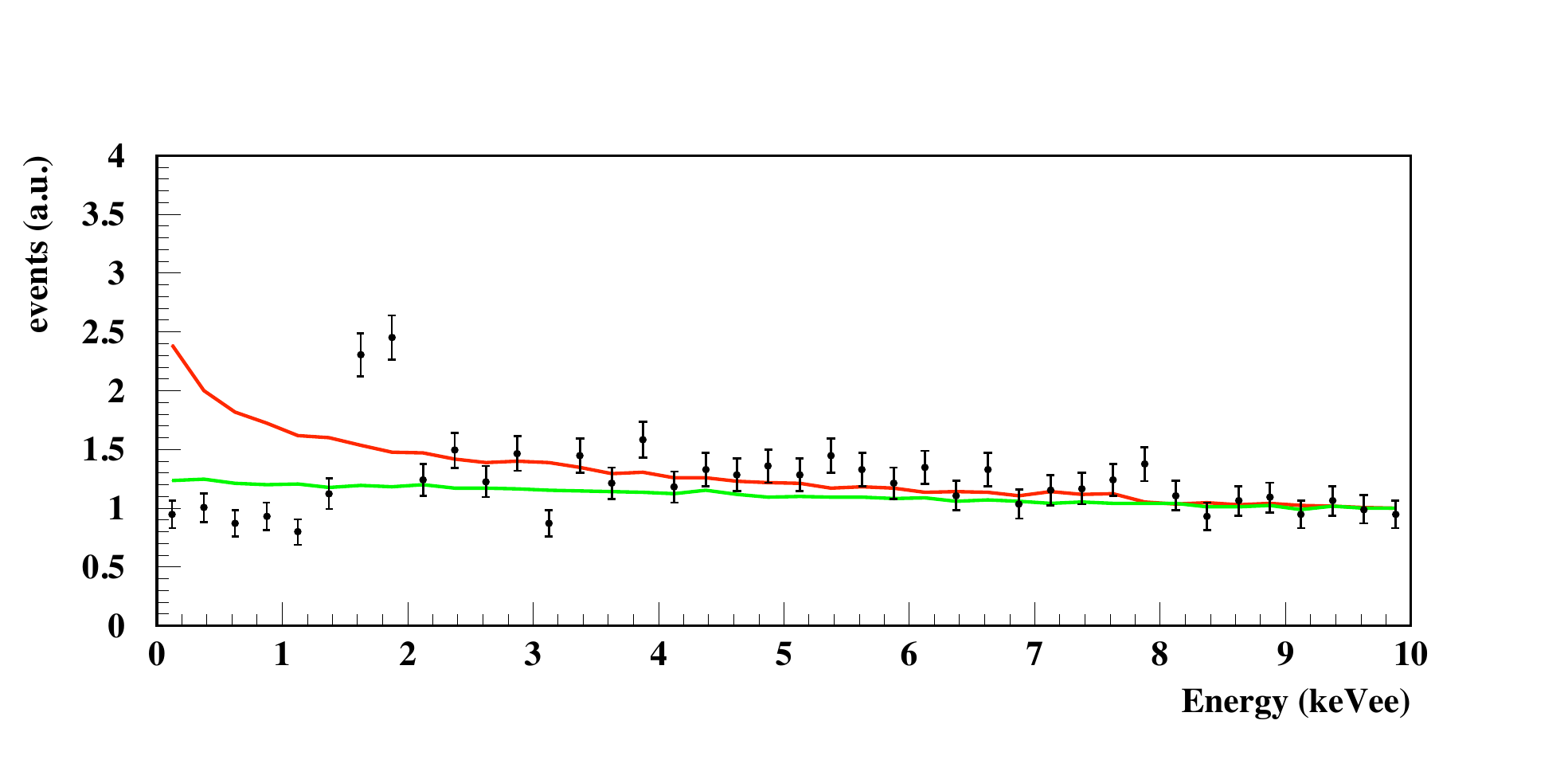}
\end{center}
\caption{Reconstructed electron equivalent energy spectrum for $^{252}$Cf exposures. The data is
consistent with expectations from Lindhard theory (red). The expectations for an energy independent
quenching factor are also shown for comparison (green). }
\label{fig:quenchcomp}
\end{figure}

In order to study $Q$, the DAMIC detector was exposed to a $^{252}$Cf  neutron source (see  Fig.\ref{fig:source}). 
%The source has a spectrum that peaks at around 1 MeV producing  a flat elastic recoil spectrum
%for energies below  keV.  The energy dependency in $Q$ transforms the 
%lat recoil energy spectrum  into a decaying spectrum when measured in electron equivalent units.
%Nuclear recoils from neutrons produce
Nuclear recoils produced by neutrons, similar to those expected from DM interactions,  give 
diffusion-limited  hits analogous to those of  X-rays. Thus, nuclear recoils in CCD
images are identified by selecting diffusion-limited hits. 
The expected response of the detector was calculated in three steps. First neutrons were
generated  according to the known spectrum of the $^{252}$Cf source and GEANT4\cite{geant4} 
was used to simulated the passage of these neutrons through the vacuum vessel wall housing 
the CCD detectors (~1cm Al wall). The resulting neutron energy spectrum was then used to calculate the
expected nuclear recoil in the Silicon CCD using Ref.\cite{Lewin}. Finally, using the ionization 
yield from the Lindhard theory, the expected observable energy on the CCD was calculated.
The results  of the simulation  compared with the spectrum for diffusion-limited hits
in data are shown in Fig. \ref{fig:quenchcomp}. 
The data shows a bump at $\sim$1.7 keVee consistent with Si excitations, but
deviates from Lindhard model at energies below 1.5 keV.   While the data appears to indicate
a weaker  energy dependence of the quenching factor below 1.5 keV, we use the Lindhard model 
in order to produce more conservative limits. For energies lower than 0.5 keV 
the detection efficiency for nuclear recoils has a strong energy dependence (see Fig.\ref{fig:neutroneff2})
and the comparison is no longer valid. This issue will be investigated in future work. 

In Fig.\ref{fig:quenchcomp} the behavior of the ionization efficiency 
of nuclear recoils is degenerate with the selection efficiency. The
spectrum is also contaminated with electron recoils produced from the
gammas generated in the $^{252}$Cf souce.
The comparison between data and simulation is also influenced 
by the neutron input spectrum and the simulation geometry. For these two reasons
the analysis discussed above, and summarized in Fig.\ref{fig:quenchcomp}, does not
constitute a measurement of $Q$. It should be interpreted as a comparison between
the data and Lindhard model assuming a constant detection efficiency. It is
presented here to illustrate the effect of energy dependent $Q$ in the recoil spectrum and
to motivate the need for a full calibration of nuclear recoils in Si at low energies.

\section{Selection of dark matter candidate events}

Three selection cuts are used to separate the dark matter candidates in our
images from background and noise hits. The first step  requires
the total energy deposited  to be larger than 0.04 keVee. This cut 
is mostly used to suppress the noise on the readout of the CCD detector. 
In order to select diffusion-limited hits produced in the bulk of the 
CCD and not near the front or back surfaces, we impose additional 
two selection described below.

X-rays  produce diffusion-limited hits similar to nuclear recoils in silicon, 
becoming an important background for a dark matter search.
However, unlike nuclear recoils, the low energy X-ray photons only 
penetrate a few microns into the detector,  producing charge very close 
to the back or front surfaces of the CCD. When performing a 
DM search it is convenient to reject hits produced near the front and back
surfaces of the CCD.

Hits on the front surface will have negligible diffusion because they
are produced next to the gates. Hits on the back surface have the maximum
diffusion since the path is the longest to the electrodes. The dependence of diffusion with 
depth can be clearly seen in tracks identified as cosmic ray muons  in Fig. \ref{fig:muon}. 
The diffusion measured for each hit becomes  indicative of the depth
of the hit inside the detector. 

For each hit reconstructed on the CCD image  the first order moments 
of the charge distribution along each axis  are calculated, namely $\hat{x}$ and $\hat{y}$. 
These values  are an estimation of the real position of the hit on the CCD coordinates, $x$ and $y$.
The second order moments along both axes are also calculated as
$\hat{\sigma}_x^2$ and $\hat{\sigma}_y^2$, providing an estimation for the charge spread 
$\sigma_x^2$ and $\sigma_y^2$.   These four estimators are the basis of the 
selection cuts used to eliminate the events near the front and back surface. In the 
following discussion units of pixel and pixel$^2$ are implicit when presenting
values for $\hat{x}$ and $\hat{\sigma}_x^2$.

Unfortunately, because of the pixelation of the detectors there is a non trivial relation
between the real values of $x,y,\sigma_x^2,\sigma_y^2$ and the estimators
 $\hat{x},\hat{y},\hat{\sigma}_x^2,\hat{\sigma}_y^2$ discussed above. For example, 
events close to the boundary of the pixel necessarily have  $\hat{\sigma}_x^2 > \sigma_x^2$. 
At the same time, events with small charge spread $\sigma_x^2$ tend to have $\hat{x}$ 
biased towards the center of a pixel. Also due to  pixelation, there
is a strong correlation between  $\hat{\sigma}_x^2$ and $\hat{x}$, given by 
\begin{equation}
\begin{split}
\hat{\sigma}_x^2 &\ge 0.25 - \hat{x}^2  \quad \mbox{for  }   \quad 0.0<\hat{x}<0.5  \mbox{  and} \\
\hat{\sigma}_x^2 &\ge 0.25 - (\hat{x}-1)^2  \quad \mbox{for  }    \quad 0.5<\hat{x}>1.0 \quad.
\label{eq:x_vs_sigma}
\end{split}
\end{equation}
With $\hat{x}=0.0$ corresponding to the edge of the pixel and $\hat{x}=0.5$  the center
of the pixel. A similar relation applies for the $y$-axis.

\begin{figure}
\begin{center}
\includegraphics[width=1\columnwidth]{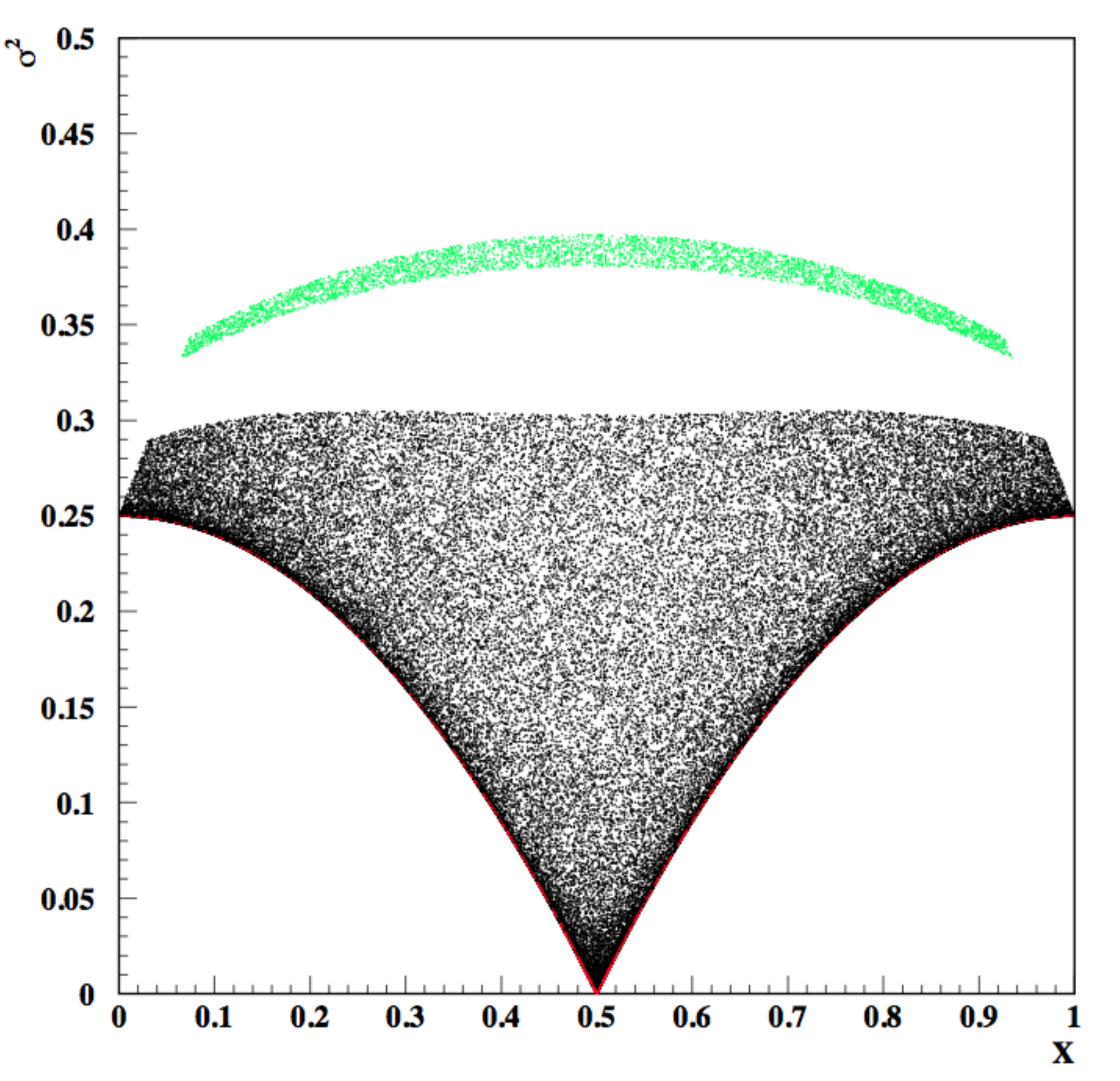}
\end{center}
\caption{Results of a simulation of charge diffusion inside a CCD. The horizontal axis shows the position of the hit inside the pixel, with $\hat{x}$=0.5
corresponding to the center of the pixel. The vertical axis shows the size of the reconstructed hit, $\hat{\sigma}_x^2$  in units of pixel$^2$ 
as defined in the text. The red points correspond to hits generated in the first 50 $\mu$m adjacent to the front of the detector, the green points
correspond to the hits generated less than 10 $\mu$m away to the back of the CCD. The black points
correspond to hits generated between 50 and 200  $\mu$m from the front of the detector. 
The simulation shows that  the selection criteria discussed in
Eq.(\ref{eq:cut3p1}) and Eq.(\ref{eq:cut3p2}) rejects most events next to the
front and back surfaces of the CCD. }
\label{fig:sizecut}
\end{figure}

A simulation was performed to illustrate the pixelation effects on
 $\hat{x}$ and $\hat{\sigma}_x^2$ . The simulation consisted of generating point-like
charge distributions in a 250 $\mu$m thick CCD with 15x15 $\mu$m$^2$ pixels.  Lateral diffusion 
for each hit was calculated as a function of the distance of the hit to the front surface
(maximum diffusion was 7.5 $\mu$m for hits 250 $\mu$m away from the front). 
The results shown in Fig. \ref{fig:sizecut} clearly indicate
that the larger   $\hat{\sigma}_x^2$ is measured for events on the back of the detector,
and that events on the front surface are very close to the curve defined by Eq.(\ref{eq:x_vs_sigma}).

\begin{figure}
\begin{center}
\includegraphics[width=1\columnwidth]{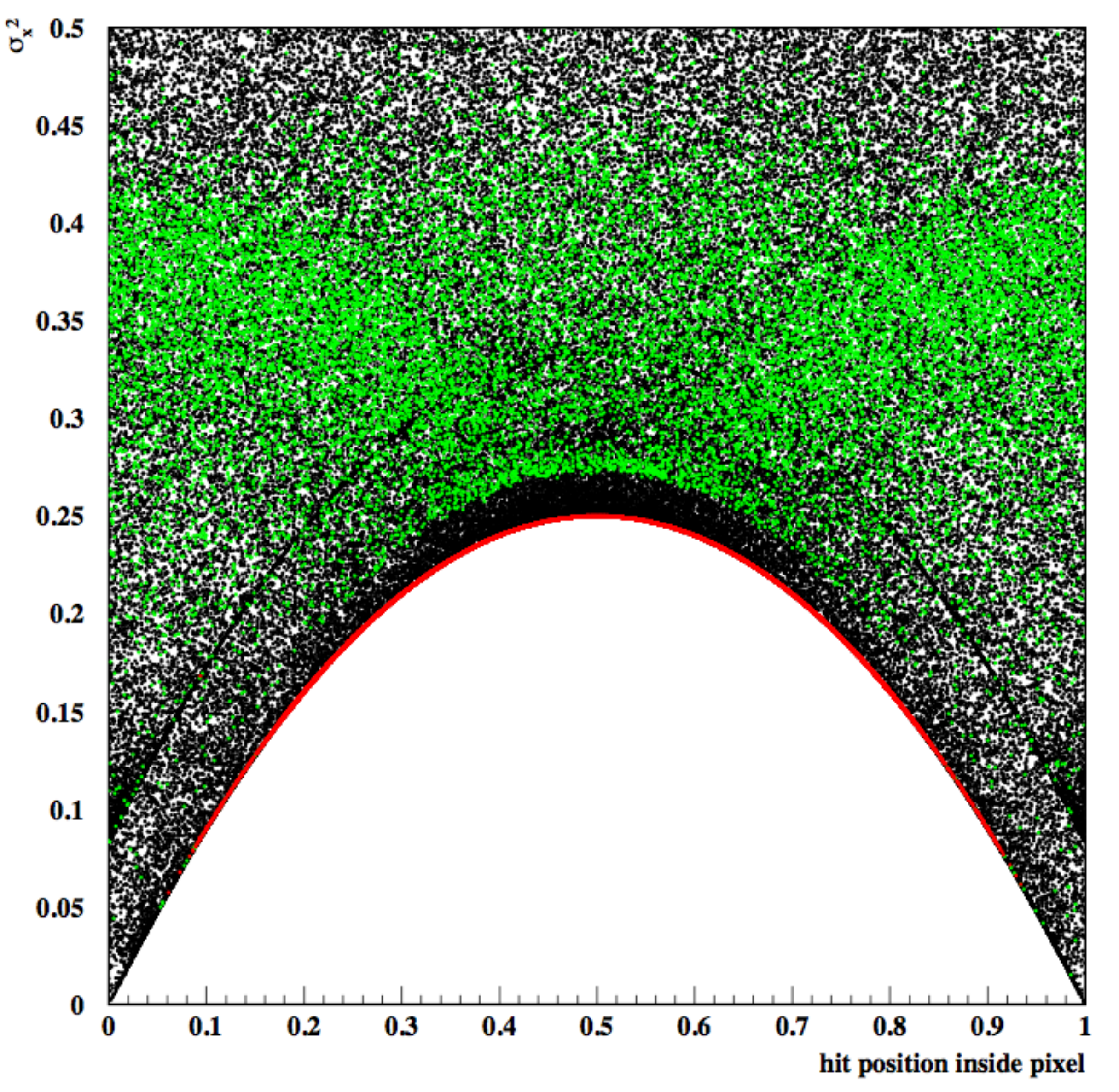}
\end{center}
\caption{Results of $\hat{x}$ and $\hat{\sigma}_x^2$ for 5.9 keV X-rays on front (red) and back (green) of a CCD. The
black data corresponds to hits from the same  $^{252}$Cf source  shown in Fig.\ref{fig:source}. As expected from simulations
the back illuminated X-rays produce hits with large  $\hat{\sigma}_x^2$, while the front illuminated X-rays are very close
to the boundary defined by Eq.\ref{eq:x_vs_sigma}. }
\label{fig:sizedata}
\end{figure}

The results for  $\hat{x}$ and $\hat{\sigma}_x^2$  with X-rays data on the front and 
back of the detector and shown  in Fig.\ref{fig:sizedata}, together with with data from a
 $^{252}$Cf  neutron source. The data shows consistency with the simulation.

Based on the discussion above, the second step for the selection of DM candidate hits consists of requiring 
the charge to be distributed in more than one pixel. This is used
to reject X-rays in the front surface of the detector.

The third selection cut is designed to keep hits in the bulk of the CCD
while rejecting events close to the front and back surfaces of the detector.
This is done in two steps, the requirement
\begin{equation}
\begin{split}
\hat{\sigma}_x^2-(0.25 -     \hat{x}^2))  > &0.05   \mbox{ for }  0.0<\hat{x} <0.5  \mbox{ and} \\
\hat{\sigma}_x^2-(0.25 - (1-\hat{x})^2) >&0.05    \mbox{ for }  0.5<\hat{x} <1.0 , \label{eq:cut3p1}
\end{split}
\end{equation}
is used to ensure that the selected events are away from the front of the detector. To select  events away from the back of the detector one additional condition is added
\begin{equation}
\hat{\sigma}_x^2 < 0.28 
\label{eq:cut3p2}
\end{equation}
Identical cuts are applied to the $y$-axis variables.

\begin{figure}
\begin{center}
\includegraphics[width=1\columnwidth]{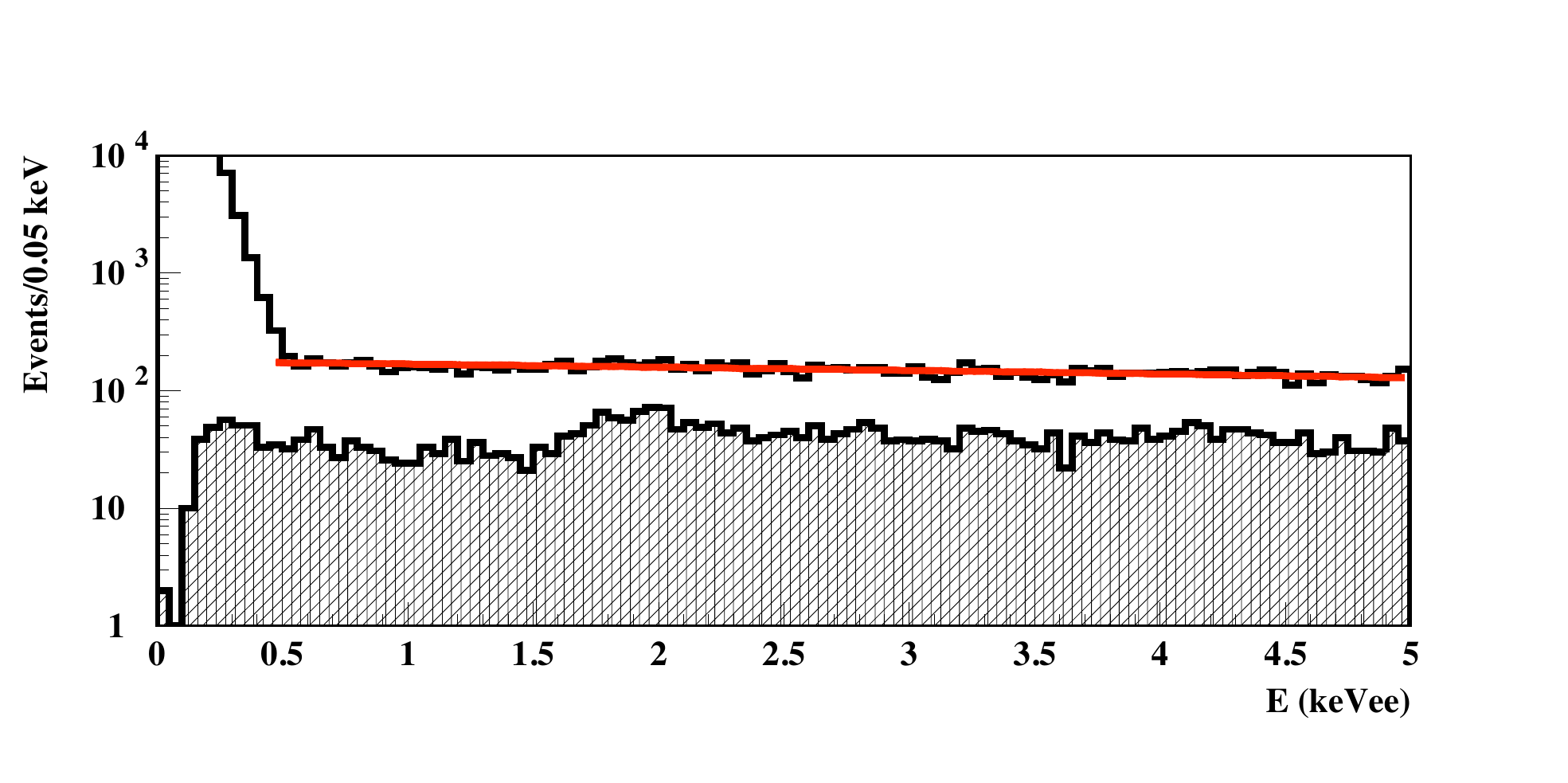}
\end{center}
\caption{ Measured spectrum for $^{252}$Cf . The upper line shown the raw data without
applying any selection criteria, the lower hatched histogram shows the result after selecting
hits with intermediate size. The raw data is fitted to a linear function above 0.5 keV. }
\label{fig:neutroneff1}
\end{figure}

\begin{figure}
\begin{center}
\includegraphics[width=1\columnwidth]{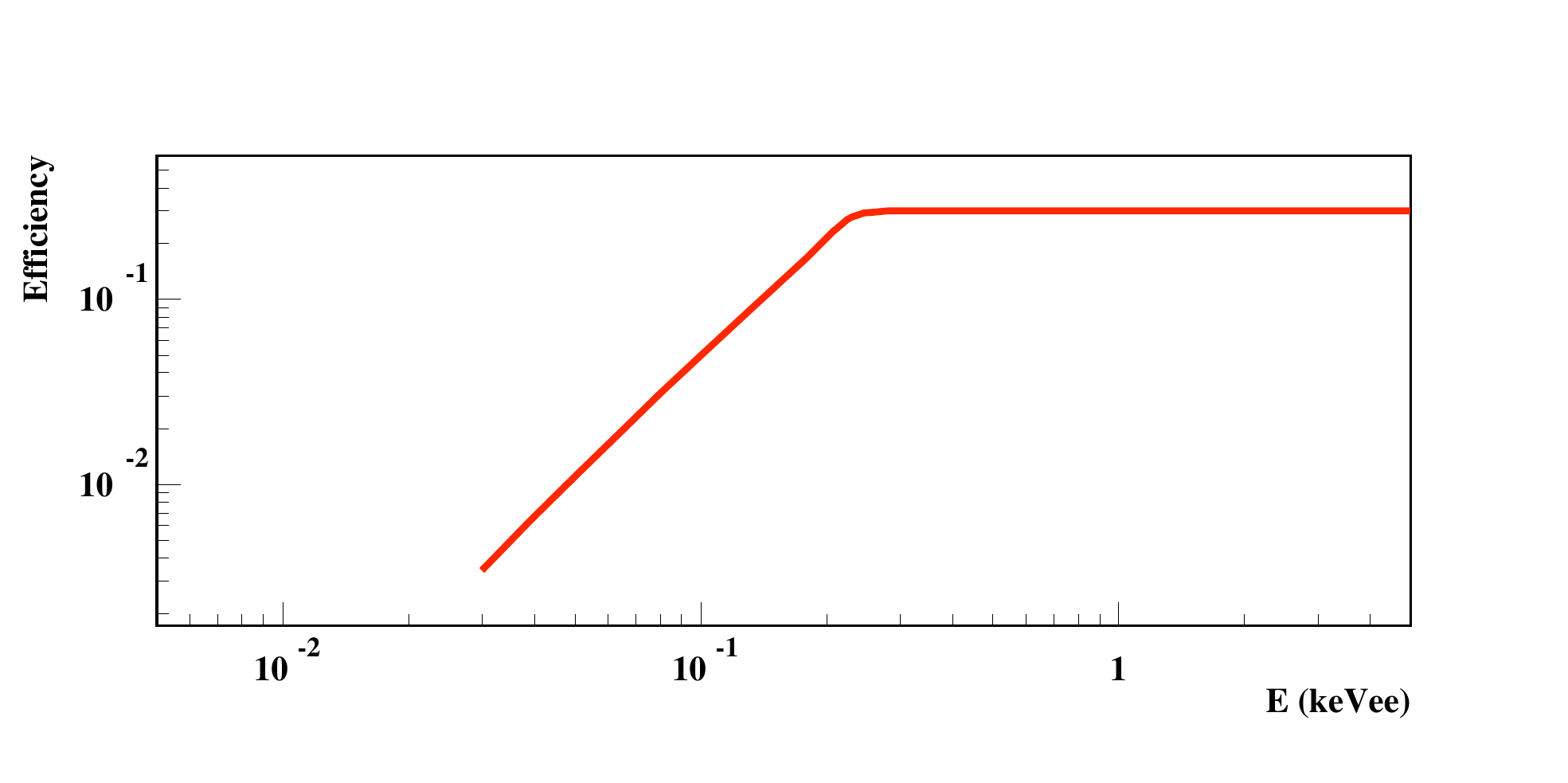}
\end{center}
\caption{Efficiency for selecting nuclear recoils with intermediate size as calculated using the data in Fig.\ref{fig:neutroneff1} }
\label{fig:neutroneff2}
\end{figure}

The efficiency for nuclear recoils passing the  selection
is estimated using  $^{252}$Cf exposures. The reconstructed
spectrum with and without the selection is shown in Fig.\ref{fig:neutroneff1}.  
The raw spectrum (without size selection) is contaminated  by low energy photons and electrons
that produce a steep raise  below 0.5 keV.
This contamination is also evident in the image shown in 
Fig.\ref{fig:source}.  Nuclear recoils from neutrons emitted by $^{252}$Cf are expected to produce an approximately flat spectrum at low energies.  We determine the efficiency of the selection by taking the ratio of the selected hits to a linear fit to the data before selection, in the region 0.5 keV to 10 keV where the spectrum is the most consistent with a flat line~(Fig. \ref{fig:neutroneff1}).   The result is shown in Fig.\ref{fig:neutroneff2}.

\section{Test in shallow underground site}

\begin{figure}
\begin{center}
\includegraphics[width=1\columnwidth]{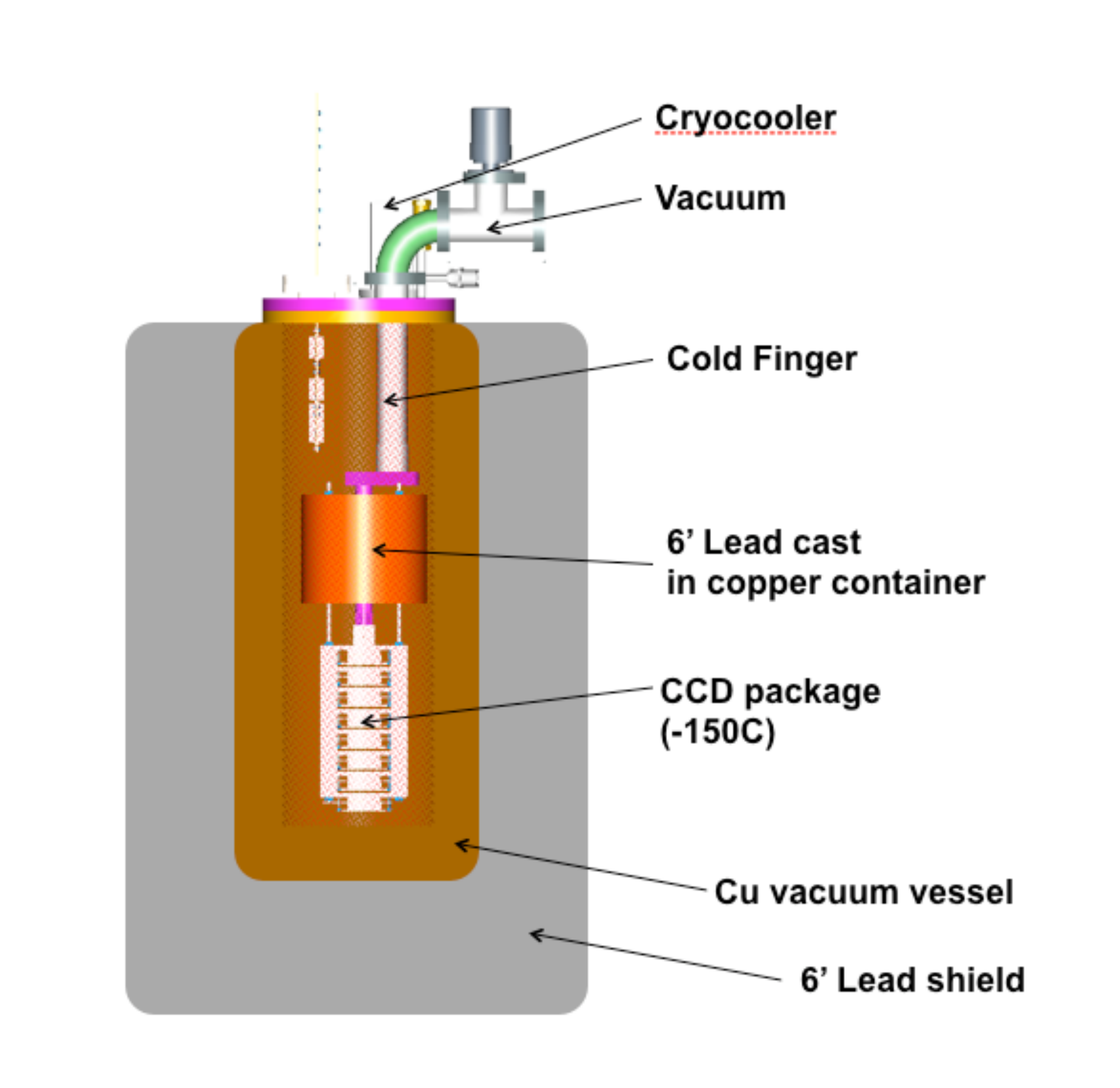}
\end{center}
\caption{Schematic of DAMIC vacuum vessel  and lead shield.}
\label{fig:dewar}
\end{figure}

We performed the first underground test for  the use
of CCDs in a direct DM search with a single 0.5 g CCD as described above. 
A low background CCD package was designed for this test,  consisting
of a readout board built on aluminum nitrade (AlN) wire-bonded to the CCD 
with some additional AlN spacers for mechanical support. AlN is typically
used in CCD packaging due to its mechanical and thermal properties and was
measured to have low concentrations of radio-isotopes. The CCD
package was installed inside a copper box 
and cooled down to -150C to reduce dark current intrinsic to the CCD. 
The copper box was also cooled to eliminate the infrared radiation impinging on the detector
from warm surfaces. A closed cycle helium gas refrigerator  
\cite{cryomech} was used to maintain the low temperature. 
 The detector was connected with a passive readout cable and  the active electronic 
components (preamplifiers) were located outside a 6'' lead shield. 
The detector package was housed in a cylindrical vacuum vessel fabricated with 
oxygen-free copper, and maintained at 10$^{-7}$ Torr by continuously running 
a turbo molecular pump as in Fig.\ref{fig:dewar}. The apparatus was installed 350' 
underground in the NuMI \cite{minos} near-detector hall at Fermilab, where other 
DM experiments have already been performed \cite{coupp}.
The system operates without the need of any human intervention, 
and was programmed to collect exposures every 40,000 seconds.
The experiment operated for  11 months starting June 2010, and accumulating a total exposure of  107 g-days. 
The observed spectrum in shown in Fig. \ref{fig:spectrum1}.  The spectrum shows several X-ray peaks
coming from known components the apparatus. It also shows three peaks at 11 keV, 12 keV and 14 keV for
which the source has not been identified. The isotopic composition of the commercial cable and 
high density connector used for the CCD have not been fully characterized, and are the most likely source
of these unidentified features. These X-ray peaks limit the DAMIC reach in a DM search, and point
to the need of improving the isotopic control of the apparatus. 

The low energy nuclear recoil
candidates were selected as described above and the resulting spectrum is shown in   Fig. \ref{fig:spectrum2}.
Given the shallow depth of the underground site, we expect that most of the events in the spectrum correspond
to neutrons produced from cosmic ray muons hitting the rock or the lead shield.  Because of the lack of timing information from
CCDs due to the long exposure, DAMIC cannot benefit from an active muon veto as is commonly used in other
DM searches.  The number of events passing each selection criteria are shown in Table \ref{tab:evts}

\begin{table}
\begin{center}
    \caption{Number of events passing selection cuts for the 107 g-day of data shown in Fig.\ref{fig:spectrum1}. two energy bins are shown:  (1) from 0.04 keV to 5 keV , (2) from 0.04 keV to 15 keV.  }
    \begin{tabular}{ | l | l | l | l |}
    \hline
     cut               &   bin (1)  &    bin (2)  \\ \hline
     1) E $>$ 0.04 keV &  81754             &   102469           \\ \hline
     2) npixel $>$ 1   &  26971             &   45353              \\ \hline
     3) hit size        &  433                  &   5529                \\ \hline
    \hline
    \end{tabular}
    \label{tab:evts}
\end{center}
\end{table}

\section{Results and Conclusion}

Standard techniques described in Ref. \cite{Lewin} were used to interpret these results as a cross section 
limit for spin-independent DM interactions. We assume a local WIMP density of 0.3 GeV/cm, dispersion velocity for the halo of
230 km/s, earth velocity of 244 km/s and escape velocity of 650 km/sec.
The Lindhard model  was used to obtain recoil energies as discussed above. 
Other possible combinations of parameters could have been used, 
in particular, the choice of escape velocity might be considered optimistic 
for this analysis. However, these choices were 
made to compare directly with  recent publications from experiments looking for
low mass dark matter particles  \cite{cogent}.

\begin{figure}
\begin{center}
\includegraphics[width=1\columnwidth]{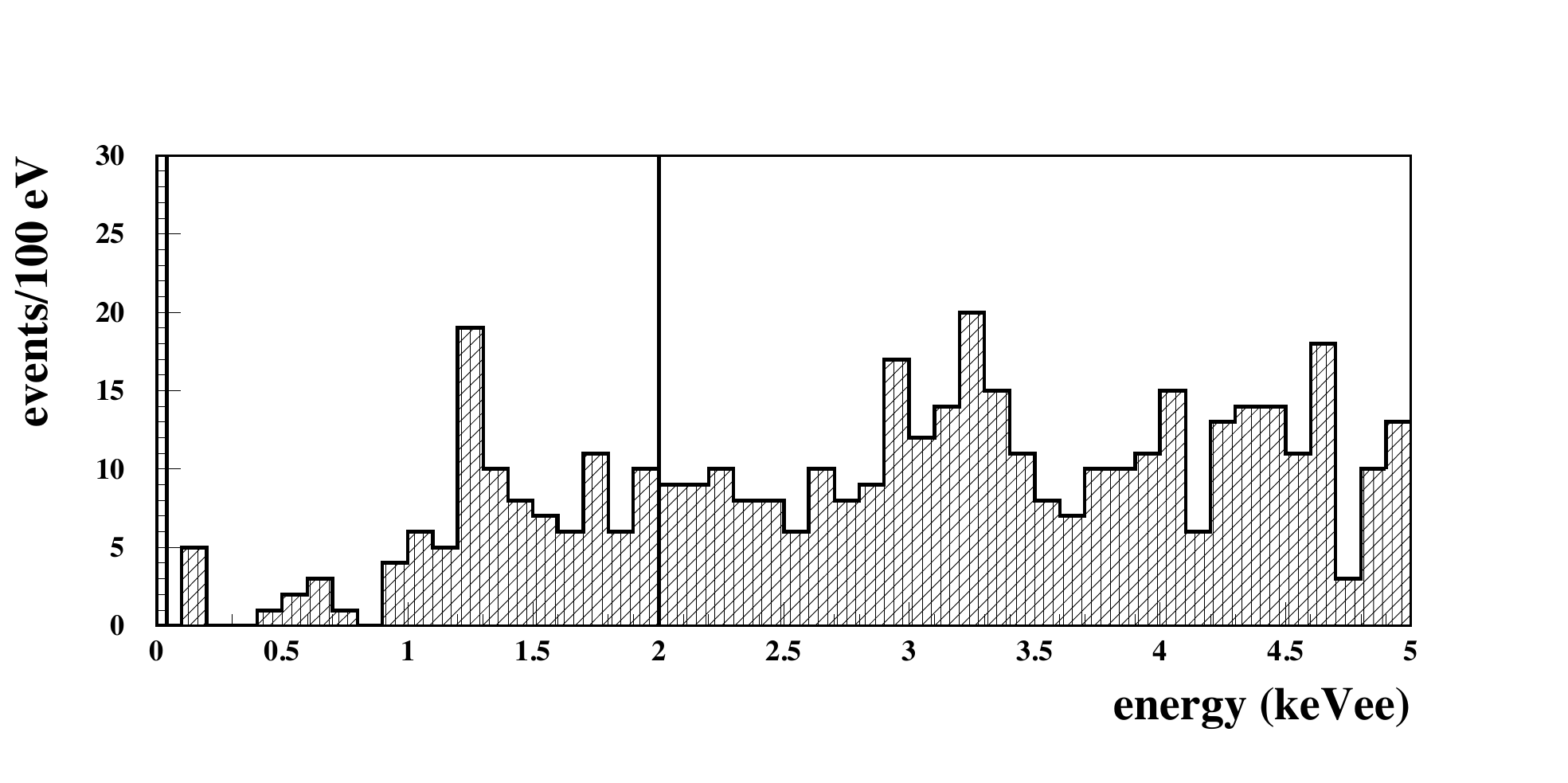}
\end{center}
\caption{Low energy spectrum of events passing the selection cuts. The vertical lines at 0.04 keV and 2 keV show the
energy range used for the DM analysis. }
\label{fig:spectrum2}
\end{figure}

\begin{figure}
\begin{center}
\includegraphics[width=1\columnwidth]{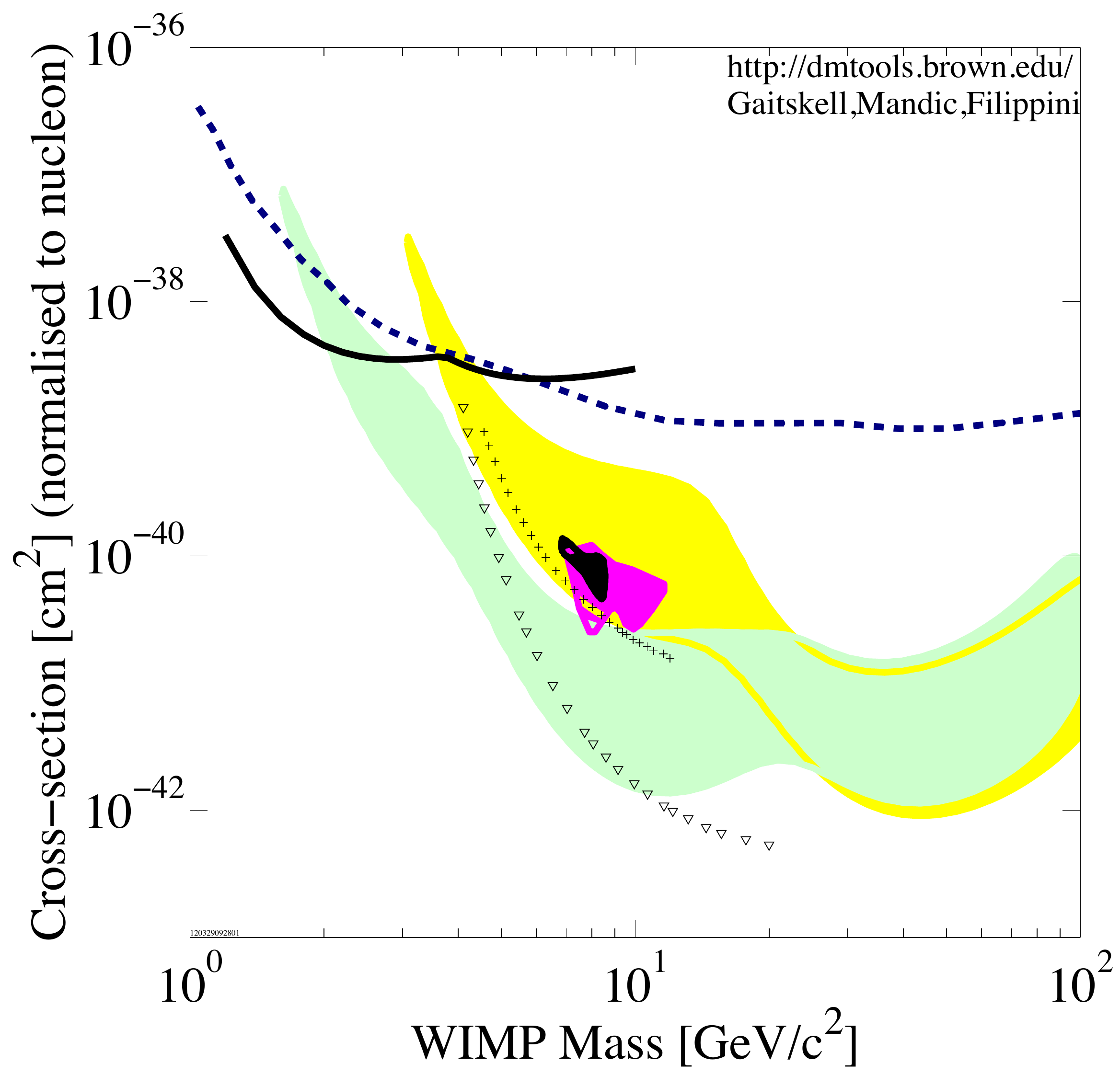}
\end{center}
\caption{ Cross section upper limit at  90$\%$ C.L. for the DAMIC results (solid black)  compared
to CRESST 2001 (dashed blue), XENON10 \cite{xenon_10} (triangles) and CDMS \cite{cdms_lowmass}  (crosses). The shaded areas correspond
to the 5-sigma contour consistent with the DAMA/LIBRA annual modulation signal 
(yellow: no ion channeling, green: ion channeling) \cite{damalibra_contour}.
The magenta contour corresponds to the DM interpretation of the CoGent observed excess and
the black contour is the region of interest for the CoGent annual modulation signal  \cite{cogent}.}
\label{fig:limit}
\end{figure}

The  optimal interval method \cite{optimal_i1,optimal_i2,optimal_i3,optimal_i4} was used for
determining the upper limit on DM cross section as a function of mass.  The energy range used
for this analysis is between 0.05 keVee and 2 keVee as shown in Fig.\ref{fig:spectrum2}.
The resulting 90$\%$ C.L. limits are shown in Fig. \ref{fig:limit}, and constitute the new 
best limit for dark matter particles of  masses below 4 GeV.  This corresponds to an improvement over the 
limits produced by  CRESST-I  \cite{cresst-2001}, which were achieved using sapphire cryogenic
detectors with a threshold of 600 eV and exposure of 1.5 kg-day.  
The region consistent with a DM interpretation of the DAMA/LIBRA signal  is shown as a 
shaded area in Fig. \ref{fig:limit} , along with recent results from
the CoGeNT collaboration interpreted as a DM signal \cite{cogent,light_DM2}. 
These two results make the low mass region very interesting and
clearly in need of further studies.  

\begin{figure}
\begin{center}
\includegraphics[width=1\columnwidth]{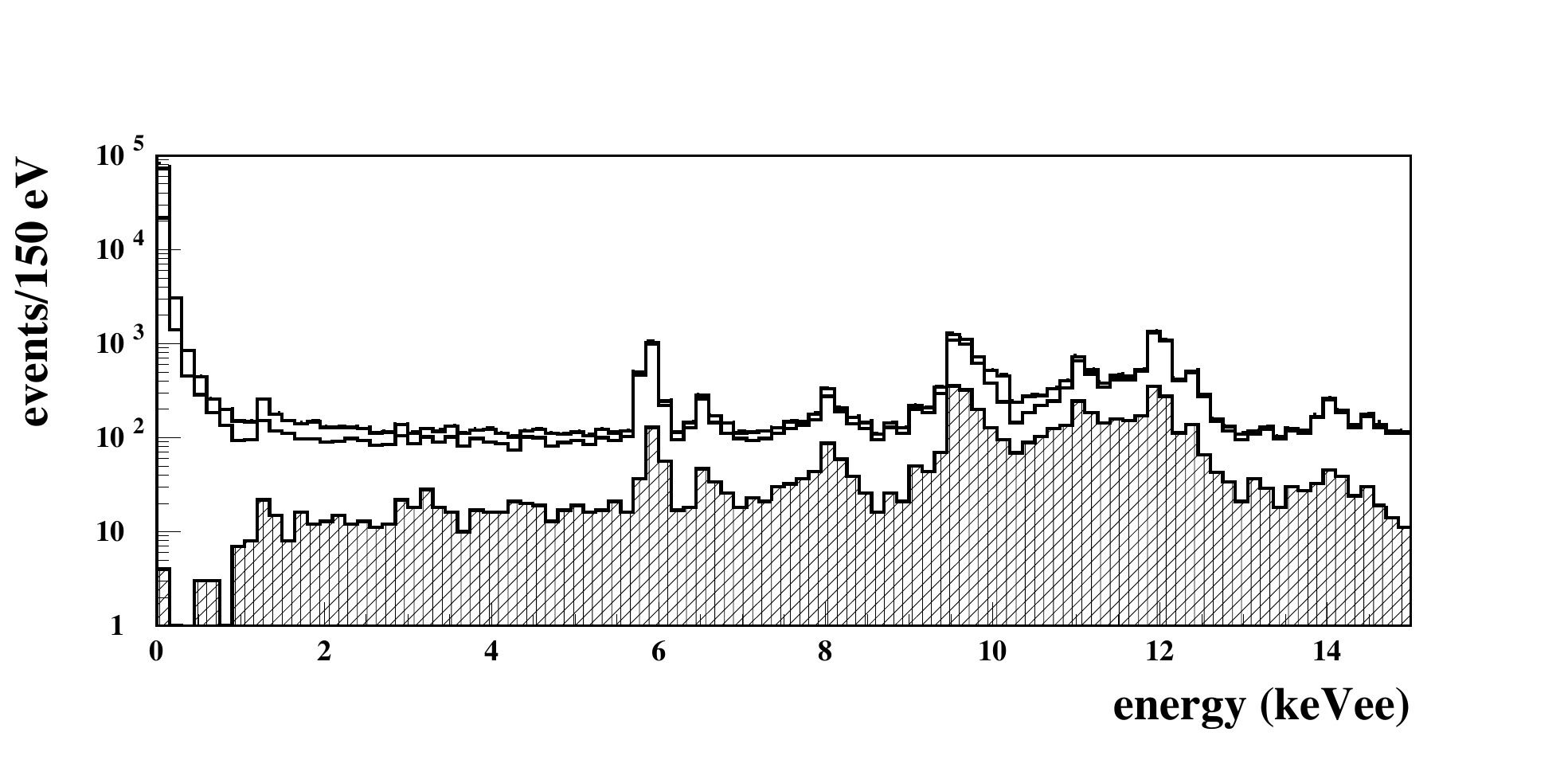}
\end{center}
\caption{Energy spectrum for nuclear recoil candidates measured in a 107 g-day exposure for DAMIC. The upper histogram
shows the spectrum for all hits, the solid line just below is the result  of eliminating the single pixel hits, 
and the hatched histogram corresponds to the events passing the selection cuts described
in the text.  The  8.0 keV peak can be attributed to Cu K$_{\alpha}$ X-rays from the material surrounding the CCD. 
The 5.9 keV and 6.4 keV peaks can be attributed to the Mn  K$_{\alpha}$ X-rays most likely from cosmogenic $^{55}$Fe. The
9.6 keV peak matches the Au L$_{\alpha}$ line. The peaks at 11 keV, 12 keV and 14.0 keV match Ge, Br and Sr  X-rays, but
these elements can not be attributed to any known component in the DAMIC apparatus.}
\label{fig:spectrum1}
\end{figure}

The test at the shallow site presented here shows an average
background rate of 600 cpd/keV (see Fig. \ref{fig:spectrum1}). This initial
setup did not have a shield for neutrons coming from the
rock. Because of the lack of timing information on CCDs, it is not
possible to take advantage of an active muon veto to remove 
cosmogenic neutrons. For these reasons the largest contribution to the
background comes from neutrons. The next step for this technology becomes then
the development of a CCD experiment in a deeper site (where
cosmogenic neutrons become negligible) and with an efficient
neutron shield.  A CCD search for DM with a background of 
10 cpd/keV and a threshold of 40 eVee would improve the limit presented here
by a factor of 60,  and could  provide a new probe for the low mass
region where hints of DM have been observed by other experiments.  
The DAMIC collaboration is currently considering an installation of the 
experiment a SNOLAB to reach this goal.

Recently we have demonstrated that 0.2 e- R.M.S. readout noise can be
achieved with a new type of CCD detector fabricated by LBNL using the same
high-resistivity Si technology \cite{pibedeoro}, but with a non-destructive output 
circuit enabling multiple readouts to reduce noise, a so-called "skipper CCD"~\cite{janesick}.
This corresponds to a 
readout noise 0.72 eV and a reduction of a factor of ten compared to the
results presented here.  A DM search with the skipper CCD would improve
significantly the reach at low energies and is being considered as part of
a DAMIC upgrade.

%The ability to perform a direct DM search with CCDs has been demonstrated using
%thick high-resistivity detectors. The low electronic readout noise for these detectors ($\sim$7.2 eV R.M.S.)
%allowed us to perform the search with a threshold of 40 eVee. Even with the extremely low
%active mass of 0.5 g, this threshold allowed us to set the best limit in searches for 
%DM particles with low mass.

%The next step for this experiment consist of
%running DAMIC in a deeper underground site with
%a larger active mass. At the same time, we are
%investigating the use of a different type of CCD that would give
%a lower readout noise (skipper CCDs \cite{janesick}) and further reduce the 
%detection threshold. We are also developing a new readout system for the CCDs,
%with digital signal processing capabilities, that provide additional filtering for
%low frequency noise and could potentially allow for a reduction
%of the detection threshold.

\section*{Acknowledgments}

We thank the Fermilab technical staff for their vital contributions, specially: Kevin Kuk, Ken Schultz, Andrew Lathrop,
Rolando Flores and Jim Tweed. This work was supported by  the U.S. Department of Energy.  We thank
to FIUNA and CONACyT in Paraguay for their support.

%We thank Juan Collar for valuable comments.

%\bibliography{arcs}

%%%%%%%%%%%%%%

%\newpage
%\appendix
%\include{planatFNAL}

\end{document}